\documentclass[published,cits]{JINST}
\usepackage{amsmath}
\usepackage[tracking,spacing,kerning,letterspace=10]{microtype}
\usepackage{textcomp}
\usepackage{booktabs}
\fussy

\title{GEM-based beam profile monitors\\ for the antiproton decelerator}
\author{S.\,Duarte Pinto$^{a}$\thanks{Corresponding author: Serge.Duarte.Pinto@cern.ch}, R.\,Jones$^a$, L.\,Ropelewski$^a$, J.\,Spanggaard$^a$, G.\,Tranquille$^a$\\
\llap{$^a$}CERN,\\ Geneva, Switzerland.\\
}

\abstract{The new beam profile measurement for the Antiproton Decelerator (\textsc{ad}) at \textsc{cern} is based on a single Gas Electron Multiplier (\textsc{gem}) with a 2D readout structure.
This detector is very light, $\sim0.4$\% X$_0$, as required by the low energy of the antiprotons, 5.3 MeV.
This overcomes the problems previously encountered with multi-wire proportional chambers (\textsc{mwpc}) for the same purpose, where beam interactions with the detector severely affect the obtained profiles.
A prototype was installed and successfully tested in late 2010, with another five detectors now installed in the \textsc{asacusa} and \textsc{aegis} beam lines.
We will provide a detailed description of the detector and discuss the results obtained.

\qquad The success of these detectors in the AD makes \textsc{gem}-based detectors likely candidates for upgrade of the beam profile monitors in all experimental areas at CERN.
The various types of \textsc{mwpc} currently in use are aging and becoming increasingly difficult to maintain.}

\keywords{gaseous detectors, gas electron multipliers, \textsc{gem}, gaseous imaging and tracking detectors}

\received{November 15, 2011}
\accepted{November 15, 2011}
\published{November 15, 2011}

\begin{document}

\section{Introduction}
\begin{figure}
\includegraphics[width=\textwidth]{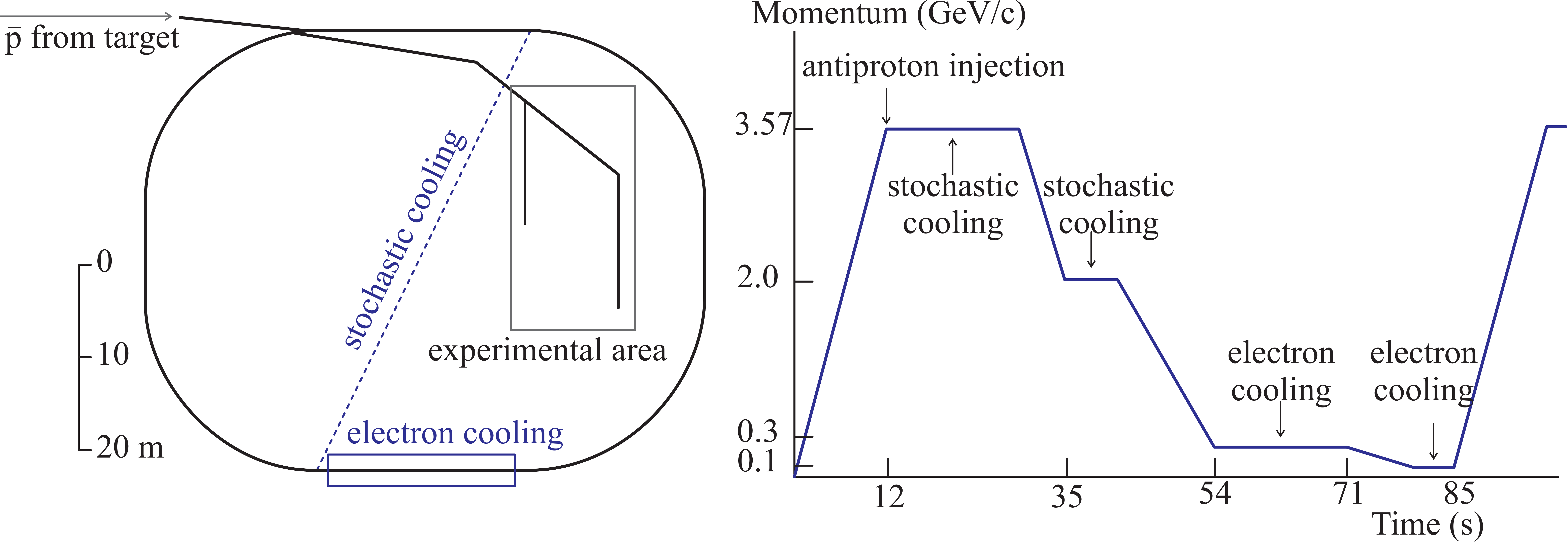}
\caption{Left: schematic view of the \textsc{cern ad}, with the experiments situated inside the decelerator ring. Right: the cooling and deceleration cycle.}
\label{AD}
\end{figure}

The antiproton decelerator~\cite{AD} at \textsc{cern} delivers antiproton beams of two different energies to five experiments, see table~\ref{experiments}.
\begin{table}[b]
\caption{Overview of the experiments on the \textsc{cern ad} beamlines and their beam requirements. The higher energy needed by the \textsc{ace} experiment greatly increases the dynamic range the beam profile detectors have to cope with.}
\center
\begin{tabular}{ll@{}rr}
\toprule
Experiment&Physics goal&Beam energy& Beam momentum\\
\midrule
\textsc{Atrap}&Antihydrogen trapping \& spectroscopy&5.3 MeV&100 MeV/c\\
\textsc{Alpha}&Antihydrogen trapping \& spectroscopy&5.3 MeV&100 MeV/c\\
\textsc{Asacusa}&Antiprotonic helium trapping \& spectroscopy&5.3 MeV&100 MeV/c\\
\textsc{Aegis}&Antihydrogen \& gravity&5.3 MeV&100 MeV/c\\
\textsc{Ace}&Antiprotons for cancer therapy&126 MeV&502 MeV/c\\
\bottomrule
\end{tabular}
\label{experiments}
\end{table}
The 3.57 GeV/c antiproton beam from the primary target enters the \textsc{ad} and undergoes a series of cooling and deceleration stages.
Figure~\ref{AD} shows schematically the \textsc{ad} setup and how the deceleration cycle is built up.
The beam is extracted to one of the experiments located inside the \textsc{ad} ring.
A spill is a few hundred na\-no\-se\-conds long and contains about $3\cdot10^7$ antiprotons.
The leng\-thy deceleration sequence imposes a delay of almost two minutes between spills.

Transverse profile information is needed at several locations along the extraction lines, to optimize the transfer optics.
It is very difficult to obtain low energy profiles without ab\-sorbing a significant fraction of the beam.
Therefore, the detector is installed in a pendulum that can move through the beam vacuum (see Fig.~\ref{pendulum}).
The inside of the pendulum is in contact with ambient atmosphere; thin stainless steel windows separate it from the vacuum.
When the pendulum is positioned in the beam, the beam is entirely absorbed while measuring a profile.
When experiments are taking data, all pendulums are in the garage position (as in Fig.~\ref{pendulum}) and the beam traverses an uninterrupted vacuum.
The tube that connects the inside of the pendulum with the outside atmosphere also serves as feed-trough for signal lines, high voltage cables, gas pipes, and a compressed air inlet that is used to cool the detector during a vacuum bake-out.
This tube is connected to a vacuum flange via a flexible bellow, and this is how the pendulum can swing in and out without causing any leaks in the beam vacuum.
\FIGURE{\includegraphics[width=.58\textwidth]{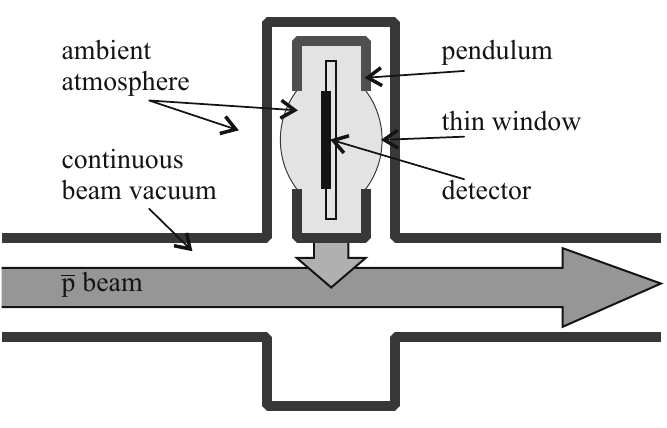}
\caption{Situation of a profile detector in a pendulum.}
\label{pendulum}}
Installation of a detector in the beam is a rather involving and lengthy operation, which requires coordination with the users of the beamline and people responsible for the beam vacuum.
First the detector needs to be installed in a pendulum, this can be done in a lab or clean room.
Then the pendulum is installed in a beam line, of which the vacuum vessel must be opened.
Due to this involving procedure there is no access to a detector once it is installed.

\section{The Detector}
\begin{figure}
\includegraphics[width=\textwidth]{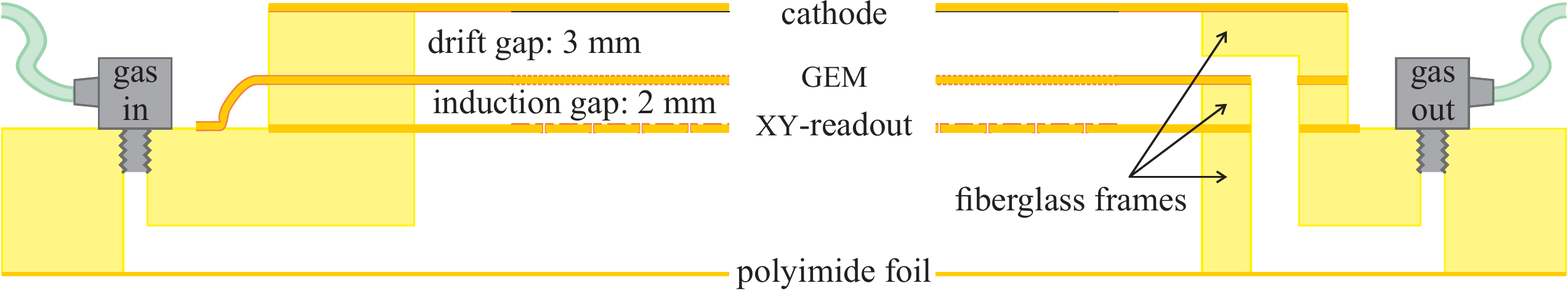}
\caption{Schematic buildup of the detector, showing components, gas features and dimensions.}
\label{detector}
\end{figure}

The detectors installed in the pendulums are gaseous radiation detectors based on a single gas electron multiplier (\textsc{gem})~\cite{firstGEM}.
The active area is $10\times10$ cm$^2$.
Figure~\ref{detector} shows schematically a cross-section of a detector; the \textsc{gem} foil is indicated between the cathode and the readout pattern.
The design of these chambers is optimized for low material budget, as the low energy beam is easily affected by multiple scattering.
Even materials downstream the active volume are kept light to minimize the effect of \emph{backsplash} from antiproton-nucleus annihilations on the measured profiles.
The whole detector presents about 0.40\%~X$_0$ of material to the incoming beam.
The upstream vacuum window of the pendulum is located about 3 cm away from the detector, and adds 0.12\%~X$_0$.

\subsection{The \textsc{gem}s}
The detector uses the most conventional size \textsc{gem}s available from the \textsc{cern} store, $10\times10$ cm$^2$.
The \textsc{gem}s are fabricated with the novel \emph{single-mask technique}~\cite{LargeGEM, LargeGEM2, LargeGEM3}.
This allows cheaper fabrication of larger foils with several \textsc{gem}s per foil, which in turn facilitates assembly.
For this application gains up to a few hundred are sufficient, so a single \textsc{gem} is used.
The material of one \textsc{gem} foil contributes $\sim 0.067$\%~X$_0$ to the material budget of the detector.

\subsection{Readout Circuit}
\noindent In the multiwire proportional chambers (\textsc{mwpc}s) used until now for measuring profiles, horizontal and vertical profiles are read out by separate chambers.
The chamber upstream causes so much multiple scattering to the beam that the profile measured by the down\-stream chamber is strong\-ly degraded.
To avoid this situation with our \textsc{gem} chambers, we designed a readout circuit to read both projections in the same plane, see Fig~\ref{readout}.

\FIGURE[l]{\includegraphics[width=.58\textwidth]{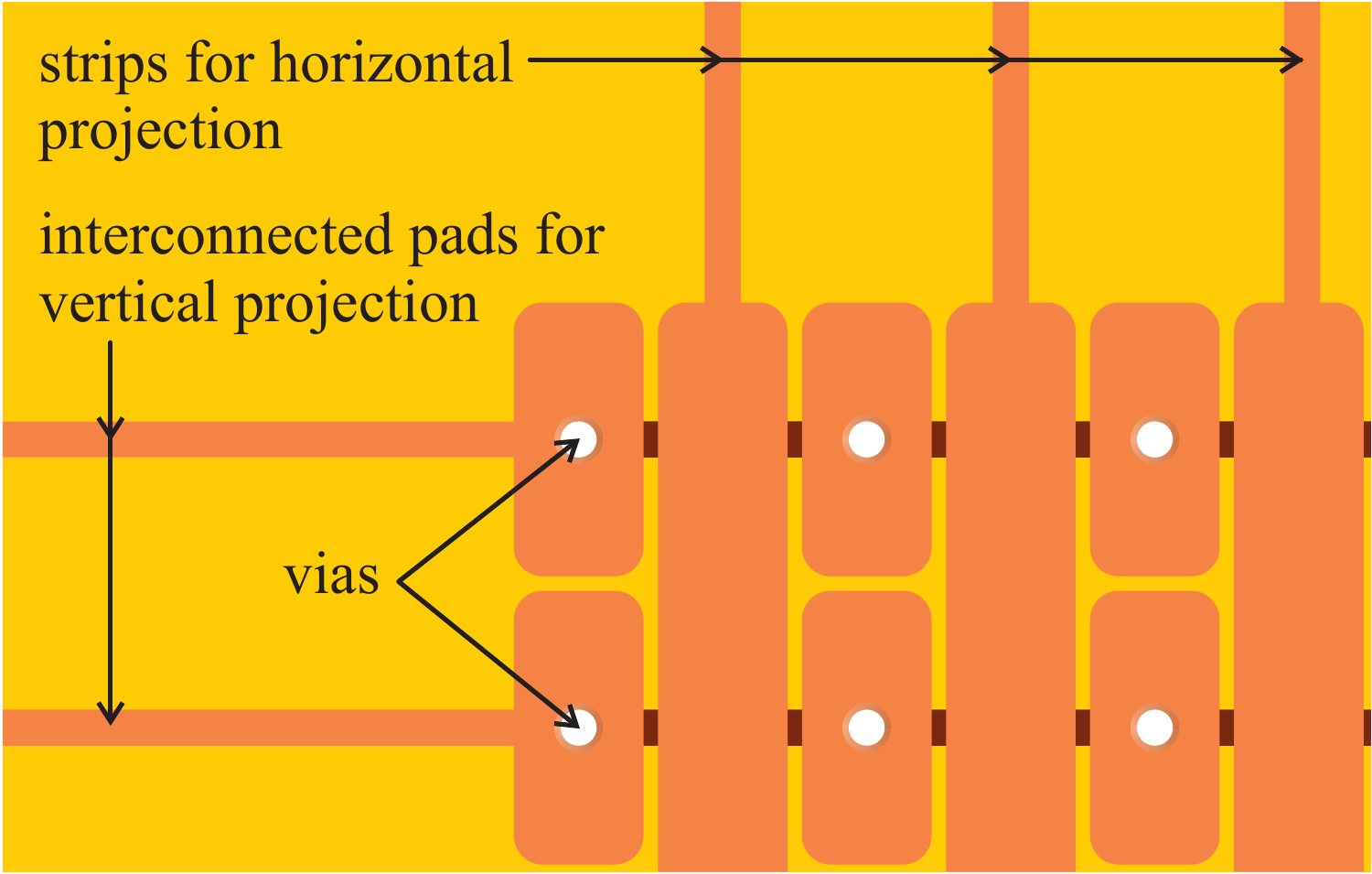}
\caption{Layout of the bidirectional readout circuit. The pitch is 1.6 mm, both horizontally and vertically.}
\label{readout}}
It is a conventional double layer\linebreak prin\-ted circuit board, with the readout elements on the top layer and signal routing traces on the bottom layer.
One projection of the profile is read out by strips, the other by rows of pads that are interconnected by traces on the bottom layer.
The collected charge is shared evenly between horizontal and vertical readout elements, and the channel density is the same horizontally and vertically (pitch: 1.6 mm).
This design can be im\-ple\-men\-ted on any base material; we chose to use polyimide in order to minimize the material budget.
The readout circuit is nevertheless the heaviest component of the detector, contributing $\sim 0.3\% \text{X}_0$.
Many techniques exist to make the many vias in such a design gas-tight, but we deliberately left them open so that gas can flow through.

\pagebreak
\subsection{Gas Distribution and Thin Cathodes}
\FIGURE{\includegraphics[width=.4\textwidth]{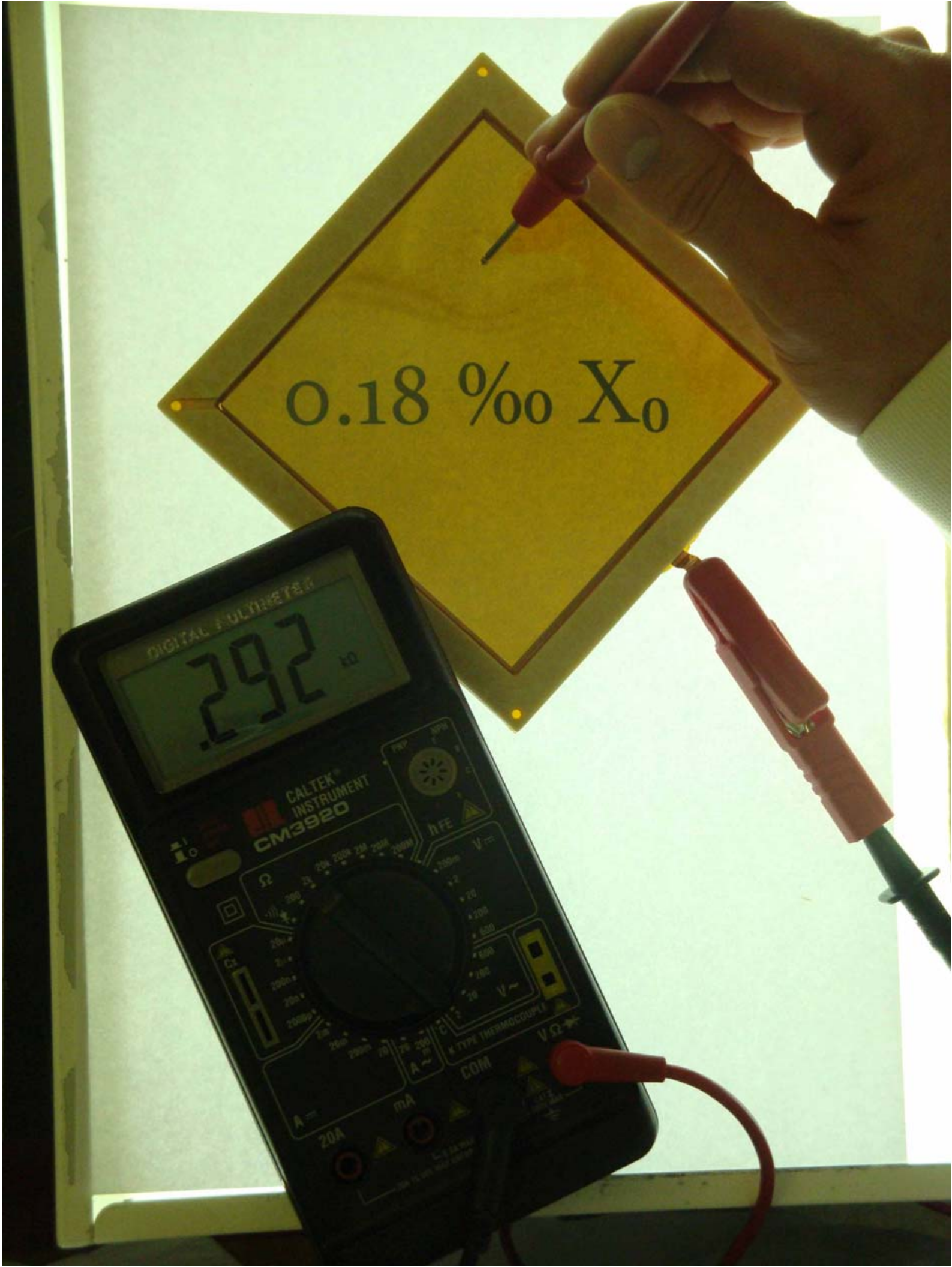}
\caption{A cathode foil stretched and glued to a spacer. The metallic layer is so thin that the foil is translucent. The resistance measured as shown ($\sim 300\,\Omega$) is reproducible and is not altered by stretching.}
\label{cathode}}
\noindent The distribution of gas trough the chamber is done by grooves milled in the fiberglass frames.
The gas flow is routed such that gas enters the chamber below the readout board, flows through the vias in the readout board and through the \textsc{gem} holes, and exits from the drift region.
This is indicated in Fig.~\ref{detector}.
The gas is a mixture of 67\% Ar and 33\% CO$_2$.

The cathode is a crucial element when absorption and multiple scattering of the beam are of concern.
In our design the cathode is also the gas enclosure, and it is stretched tight ($\sim 11$ MPa) in order to avoid any deformation by the slight overpressure in the chamber.
It is made of the same base material \textsc{gem}s are made of: copperclad polyimide.
The copper is removed in the active area of the detector, leaving just a thin ($\sim100$ nm) layer of chromium which is there to act as a tie coat for a better adhesion of the copper layer to the polyimide substrate.
The material traversed by the beam to enter the active volume of the detector thus amounts to 0.018\%~X$_0$.
On the other end of the chamber, the gas enclosure is made of a 25 \textmu m polyimide foil, adding 0.009\%~X$_0$.

\subsection{Electronics}
The electronics used to read out the detector comes from the design used in other experimental areas at \textsc{cern}, but some modifications were necessary.
It is based on a conventional switched integrator circuit built around the IVC102 unit from Texas Instruments, see Fig.~\ref{integrator}.
\FIGURE[l]{\includegraphics[width=.58\textwidth]{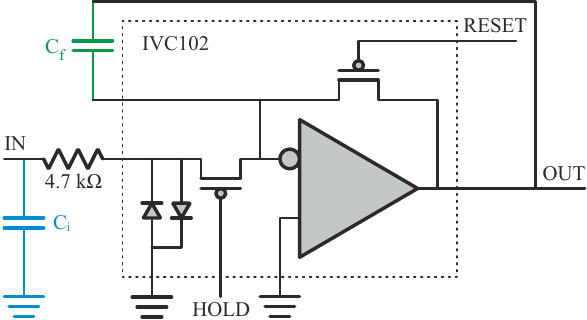}
\caption{Switched integrator circuit based on the IVC102.}
\label{integrator}}
After integrating during 20 ms, amplitudes from 64 of such integrators (32 for each projection) are multiplexed and converted by an \textsc{adc}.
More details about the acquisition system are given in~\cite{Jens}.

\begin{figure}
\center\includegraphics[width=.8\textwidth]{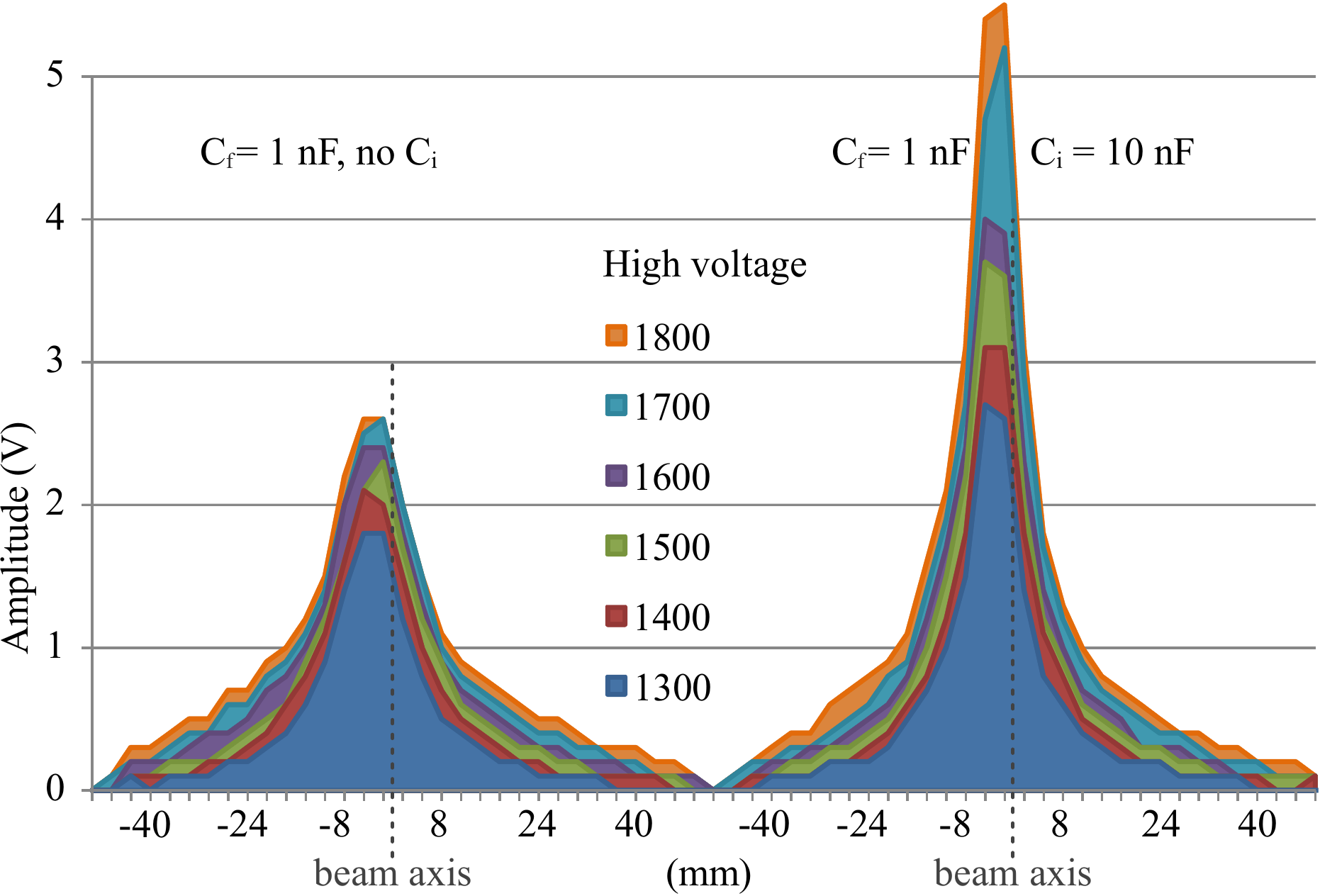}
\caption{Low energy ($E_\textrm{kin}=5.3$ MeV) beam profiles made with integrators without (left) and with (right) 10 nF input capacitors. The profiles are taken using a range of high voltage settings. The voltage indicated in the legend is the voltage applied to a resistive divider and corresponds to the voltage on the cathode. The voltage over the \textsc{gem} is always $0.22\times$ the cathode voltage.}
\label{profiles}
\end{figure}

The integrator originally had a large feedback capacitance (C$_\textrm{f}=1$ nF) in order to have a low sensitivity to noise and cross-talk.
This low sensitivity allowed the integrators to be located outside the pendulum, about two meters of cable away from the detector.
Another consequence of the low sensitivity is that a lot of charge needs to be collected to reach the full scale of the \textsc{adc} that reads out the integrator ($V_\textrm{fs}=\pm 10$~V): $V_\textrm{fs}\cdot C_\textrm{f} = 10\cdot10^{-9} = 10$~nC per channel.
This charge is collected during a spill of a few hundred ns duration, giving rise to input currents of up to 100 mA.
The protection diodes indicated in Fig.~\ref{integrator} start clamping from an input current of 0.2 mA, because of the voltage drop over the series resistance of the \textsc{fet} switch marked \textsc{hold} ($\sim1.5$ k$\Omega$).
To limit the input current and avoid the non-linear behavior caused by the clamping diodes we added a capacitor to the input of each channel (C$_\textrm{i}=10$ nF), indicated in blue in Fig.~\ref{integrator}.
This capacitor acts as a low-pass filter, collecting charge during a spill and then dissipating it slowly through the resistive elements into the integrator.
The effect of this modification can be seen in Fig.~\ref{profiles}, where profiles made with integrators without (left) and with (right) an input capacitor are shown for a range of high voltage settings.
At low amplitude there is no appreciable difference, but at high amplitude the modified integrators collect more than twice as much charge.
This is an easy way to use rather slow integrating electronics with beams with a fast spill structure.

Still, also after this modification profiles get somewhat distorted at higher voltages.
While the amplitude in the tails increases exponentially with the voltage applied (as one should expect), the center of the profile seems to saturate.
We attribute this to a saturation of the \textsc{gem} gain due to the high space charge density.
Measurements done with an ionization chamber confirm that the space charge density in the center of the profile is of the order $10^{12}$ e$^-$/cm$^3$, a density normally only seen inside an electron avalanche.
This is again a consequence of the fast spills of the \textsc{ad}, and of the low energy of the antiprotons ($\sim 3$ MeV after traversing the stainless steel vacuum window).
This issue has been solved by increasing the sensitivity of the integrators tenfold (C$_\textrm{f}=100$ pF, indicated in green in figure~\ref{integrator}), so that no gain is required from the \textsc{gem}.
A voltage below 1000 V is applied to the chamber corresponding to a \textsc{gem} voltage of $\sim200$ V, where the \textsc{gem} is \emph{transparent} to drifting electrons without multiplication.
The \textsc{gem} is still needed to deliver a gain of $\sim 100$ in case a 126 MeV beam is used; at this energy there is no issue with space charge.

\section{Conclusions \& outlook}
\FIGURE{\includegraphics[width=.56\textwidth]{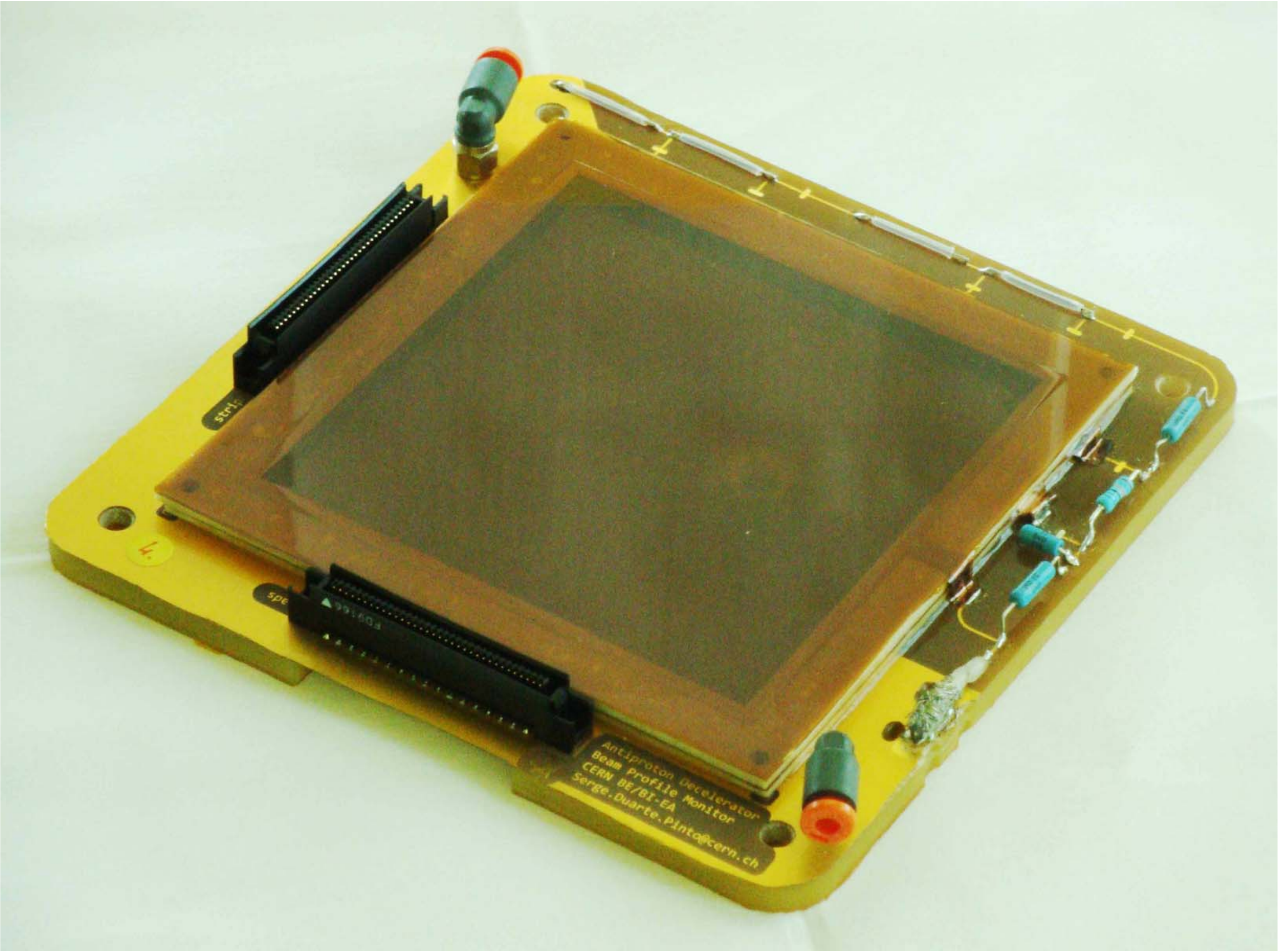}
\caption{A single \textsc{gem} detector before installation.}
\label{proto}}
\noindent We developed new transverse beam profile monitors for the \textsc{cern ad} beam lines, and report on the first results with these detectors.
The monitors are single \textsc{gem} detectors with a bidirectional readout structure, and the chamber is designed to minimize material budget.
They dramatically improve the profile measurements compared to the \textsc{mwpc}s currently used.

The electronics and the mode of operation of the detector have been optimized to accommodate the beam properties foreseen for the \textsc{ad}.
Detector response at both beam energies can be adjusted to the dynamic range of the readout electronics without causing any distortion.

At the time of writing the detectors described above (figure~\ref{proto}) are installed in 6 locations in the \textsc{asacusa} and \textsc{aegis} beam lines, and replacement of the remaining \textsc{mwpc} monitors in all beam lines is foreseen for the shutdown of 2011/2012.
All these detectors give reliable, useful profiles for operators to steer the beam.

These detectors are very easy to operate and maintain, and based on inexpensive components.
This makes detectors of a similar design likely candidates for replacement of \textsc{mwpc}s currently in use in many high energy beam lines at \textsc{cern}.

\end{document}